\documentclass{article}
\usepackage{dcase2020,amsmath,graphicx,url,times,booktabs, tabularx}
\usepackage{enumitem}
\usepackage{float}
\usepackage{multirow}
\usepackage{booktabs}
\usepackage[table,xcdraw]{xcolor}
\usepackage{soul}
\usepackage{mathtools}
\usepackage{textcomp} 
\title{Listen carefully and tell: an audio captioning system based on residual learning and gammatone audio representation}

\name{Sergi Perez-Castanos$^{}$,
    Javier Naranjo-Alcazar$^{1}$,
      Pedro Zuccarello$^{}$, 
      Maximo Cobos$^{1}$
      }

\address{         
        $^1$ Universitat de Val\`encia, Burjassot, Spain, \{janal2@alumni.uv.es\}, \{maximo.cobos\}@uv.es\\
 }

\begin{document}
\ninept
\maketitle

\begin{sloppy}

\begin{abstract}
Automated audio captioning is machine listening task whose goal is to describe an audio using free text. An automated audio captioning system has to be implemented as it accepts an audio as input and outputs as textual description, that is, the caption of the signal. This task can be useful in many applications such as automatic content description or machine-to-machine interaction. In this work, an automatic audio captioning based on residual learning on the encoder phase is proposed. The encoder phase is implemented via different Residual Networks configurations. The decoder phase (create the caption) is run using recurrent layers plus attention mechanism. The audio representation chosen has been Gammatone. Results show that the framework proposed in this work surpass the baseline system in challenge results.
\end{abstract}


\section{Introduction}
\label{sec:intro}

\par Audio captioning is a novel machine listening task that was first presented in \cite{Drossos_2017_waspaa}. Audio captioning can be understood as a intermodal translation. Its goal is to create an autonomous and smart description on an audio signal. The state-of-the-art solution employs an encoder-decoder architecture \cite{Drossos_2020_icassp}. The encoder block embeds the audio representation (i.e. log Mel-Spectrogram representation) into a lower dimensionality feature map while the decoder creates a sequence of words from that new representation, that is, the smart caption. This caption must be as close as possible as human performance which means that captions must be structured according to the language in which it is being described. This task differs from other classic machine listening tasks such as audio tagging or sound event detection. Audio captioning is not intended to assign labels to audio or to calculate onset and offset times.

\par First approximations to autonomous captioning where done in image domain \cite{vinyals2015show, you2016image} followed by autonomous video captioning \cite{venugopalan2014translating, shin2016beyond}. In \cite{vinyals2015show}, this problem was addressed for the first time. Captioning, whether in the image or audio domain, can be interpreted as an artificial intelligence problem that connects two fields. In the case of the image domain, computer vision and language processing techniques must be merged. In audio domain, the representation of the audio, the processing of this representation (similar to computer vision techniques) and the language processing must be taken into account. The intuition on which this work is based is in the work done in sentence translation where a sentence must be translated from an initial language to a target language. The emergence of recurrent neural networks (RNNs) led to the creation of simpler solutions (without reordering or individual processing of each word) and maintained state-of-the-art performance \cite{bahdanau2014neural,cho2014learning}. In this type of problem, an RNN encodes the input sentence into a fixed-size codification and a RNN decoder generates the translated sentence in the target language. As image captioning takes as input a fixed size image, in \cite{vinyals2015show} it is decided to replace the encoder block by a convolutional neural network (CNNs). This type of network has shown good results in extracting descriptive information about the images. Thus, a network of this nature will be used to encode the images for the decoder, which in this case is a recurrent network.

\begin{figure*}[t]
  \centering
  \centerline{\includegraphics[scale=0.60]{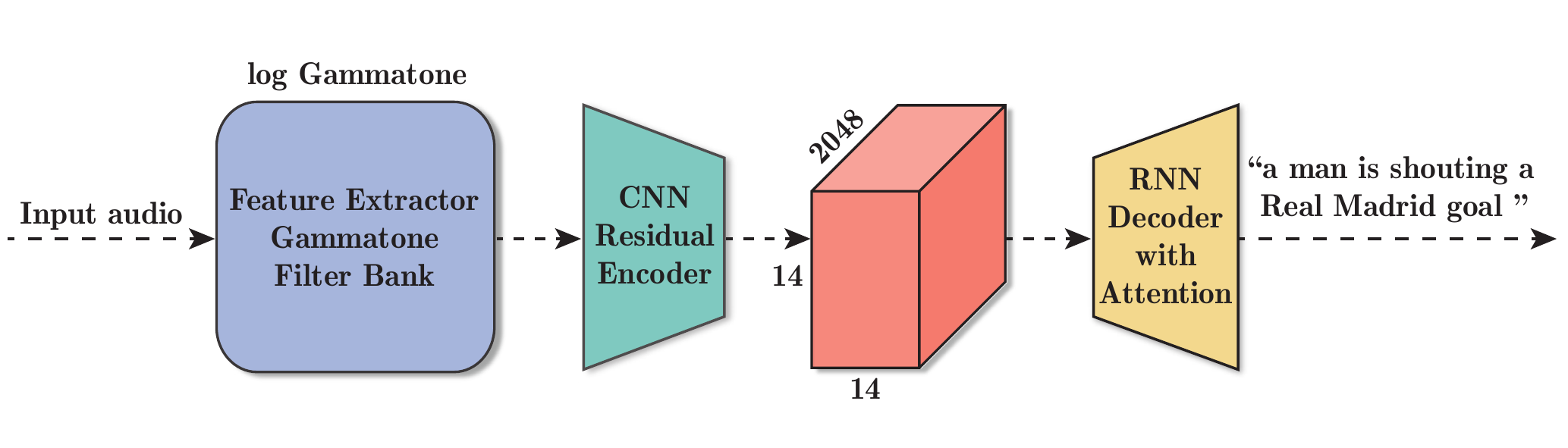}}
  \caption{Proposed system representation. Audio is first represented using Gammatone filter bank. Later, a richer representation of the audio is obtained by means of a residual encoder. The final caption is obtained by using a recurrent decoder with the attention mechanism.}
  \label{fig:framework}
\end{figure*}

Autonomous captioning frameworks (regardless the data domain, e.g. image or audio) can be divided in two blocks as mentioned before: encoder and decoder. Going a little more in detail in each of the parts can be stated:

\begin{itemize}
    \item \textbf{Encoder}: it processes the input data e.g. RGB image and creates more sophisticated high level representations from the input. This block is the one that has produced the most different solutions. Some state-of-the-art works propose the encoder to be composed of convolutional layers \cite{vinyals2015show, venugopalan2014translating}, while others by recurring layers \cite{cho2014learning, bahdanau2014neural, Drossos_2017_waspaa}.
    
    \item \textbf{Decoder}: it takes the encoder's output and creates the final caption. This block is usually implemented using a Recurrent Neural Network (RNN). The final layer is usually a fully connected one \cite{cho2014learning, bahdanau2014neural} with the number of units equal to number of possible words \cite{Drossos_2017_waspaa}. 
\end{itemize}

\par As it can be observed, an autonomous captioning system can be reformulated as a system that, given an entry $ \mathbf{X} $, the system is able to obtain the most relevant characteristics that allow it to title the entry through a series of meaningful words lexicon. 

The aim of this work is to propose a solution based on an encoder-decoder structure where the encoder corresponds to a convolutional neural network and the decoder to a recurrent neural network. To achieve more accurate results the decoder implements an attention mechanism. An analysis of different state-of-the-art residual networks as possible encoders is carried out. In addition, the representation of the audio is done using the Gammatone filter bank.

\par The rest of the paper is organized as follows:  Section~\ref{sec:method} explains the proposed method in this paper. Section~\ref{sec:exp} describes the dataset and the training procedure. Section~\ref{sec:results} shows the results obtained by all the experiments compared to the baseline. Finally, Section~\ref{sec:conclusion} concludes this work.

% Section~\ref{sec:background} provides a little insight into how data should be processed when working on the task of audio captioning.

% Below is an example of how to insert images. 
% -------------------------------------------------------------------------
% \begin{figure}[t]
%   \centering
%   \centerline{\includegraphics[width=\columnwidth]{fig1a}}
%   \caption{Example of a figure with experimental results.}
%   \label{fig:results}
% \end{figure}

% \begin{equation}
%   \label{eqn:wave_equation}
%     \Delta^2p(x,y,z,t)-
%     \displaystyle\frac{1}{c^2}\frac{\partial^2p(x,y,z,t)}{\partial t^2}=0,
% \end{equation}

% \section{Background}\label{sec:background}

% \par When training a captioning system in either the image or audio domain, it is necessary to take a few extra steps to convert the target sentences into information that can be understood by a machine learning system. A post-processing for captions is needed in order to train the system. The steps carried out are: remove all punctuation; change all letters to small case; tokenization, that is, assign specific index identification to each word; add start of sequence token, i.e. \texttt{\textbf{<sos>}} and add end of squence token and pad when necessary to the maximum caption length possible, i.e. \texttt{\textbf{<eos>}}.

\section{Method}\label{sec:method}

\par The system proposed in this work is based on \cite{vinyals2015show, xu2015show}. The main idea on \cite{xu2015show} is that the final step of the captioning system (decoder) learn where to look in the representation in order to predict the next sequence word. This mechanism is known as attention and forces the system to look for the relevant part in the encoding feature map. A full representation of the proposed framework can be seen in Figure~\ref{fig:framework}.

\begin{table*}[]
\centering
\scalebox{0.85}{
\begin{tabular}{cccccccccc}
\toprule 
\textbf{System}         & \textbf{BLEU\textsubscript{1}} & \textbf{BLEU\textsubscript{2}} & \textbf{BLEU\textsubscript{3}} & \textbf{BLEU\textsubscript{4}} &
\textbf{METEOR} &
\textbf{ROUGE\textsubscript{L}} & \textbf{Cider}  & \textbf{SPICE}  & \textbf{SPIDEr} \\
\toprule
Baseline       & 0.389   & 0.136   & 0.055   & 0.015 & 0.084  & 0.262                                   & 0.074  & 0.033  & 0.054  \\ \midrule
Enc-50-DecAtt       & 0.353   & 0.132   & 0.054   & 0.023  & 0.089 & 0.248                                   & 0.078  & 0.037  & 0.058  \\ \midrule
Enc-101-DecAtt & 0.357  & 0.138  & 0.059  & 0.026 & 0.091 & 0.249                                  & 0.088 & 0.038 & 0.063 \\ \midrule
Enc-152-DecAtt &  0.348       &   0.129      &    0.054     &   0.026    & 0.090         &                          0.246     &     0.079    &    0.038    &   0.058    \\ \midrule
   Enc-Wide101-DecAtt     &    0.345     &       0.129   &   0.054      &  0.023        & 0.090 &              0.245  & 0.079     & 0.038   &  0.058     \\ \bottomrule
\end{tabular}}
\caption{Results with different Residual configurations in development stage. Metrics obtained from evaluation subset. Teaching forcing method is used for this metrics.}
\label{tab:results}
\end{table*}

\subsection{Audio representation}\label{subsec:audio}

\par The input of this framework has been decided to be the Gammatone audio representation \cite{tabibi2017investigating, zhang2018deep}. The number of frequency bins is set to 64 as well as the baseline system. All post-processing of the captions is done with the same procedure as the baseline (see subsection~\ref{subsec:baseline}). This work implementation can be found in the following link\footnote{https://github.com/sergipc22/dcase20\_task6/tree/develop}.

\subsection{Residual Convolutional Encoder}\label{subsec:encoder}

\par In this work, the encoder block correspond to a Convolutional Neural Network (CNN). Residual networks have been the choice. In particular, residual CNNs have been the choice. Residual CNNs were first introduced in \cite{he2016deep}. Residual layers are designed in order to approximate a residual function as $\mathcal{F}(\mathbf{X}) \coloneqq \mathcal{H}(\mathbf{X})-\mathbf{X}$, where $\mathcal{H}$ represents the feature to be fit by a set of stacked layers and $\mathbf{X}$ represents the input to the first of such stacked layers. Therefore, $\mathcal{H}$ can be expressed as $\mathcal{H}(\mathbf{X})~=~\mathcal{F}(\mathbf{X}) + \mathbf{X}$. The reason why this kind of CNN have been chosen corresponds to the idea that the network training might be easier if optimizing a residual mapping instead of an unreferenced one, as in a classical CNN \cite{simonyan2014very}. This residual learning can be easily implemented by adding a shortcut connection that would perform as identity mapping, that is, adding the input  $\mathbf{X}$ to the output of the residual block $\mathcal{F}(\mathbf{X})$. In this work, different state-of-the-art Residual networks have been implemented as encoder block \cite{he2016deep, zagoruyko2016wide}.

The \textit{Resnet50}, \textit{Resnet101} and \textit{Resnet152} residual architectures were proposed in the original paper where the residual learning was first presented \cite{he2016deep}. The idea behind them is the same and they are implemented with the same convolutional blocks placed in the same sequential way. That is, they have the same kernel size or number of filters. The difference between them is how many times each convolutional block is repeated. Thus \textit{Resnet50} is a 50-layer residual network, \textit{Resnet101} is a 101-layer network, and \textit{Resnet152} is a 152-layer network.

Since the introduction of this type of network, a multitude of combinations or studies have been proposed to improve the behaviour of the residual network, either in reducing the number of layers or the training time by maintaining or improving the results obtained. One solution that aims to maintain the benefits of residual learning but with less deep networks is the Wide Residual Networks \cite{zagoruyko2016wide}. In this submission, the \textit{Resnet101} network (\textit{Wide101}) has been tested but implementing the modifications presented in \cite{zagoruyko2016wide}.

All implementations have been done with the Deep Learning Pytorch library \cite{pytorch}. It is important to point out that the networks are trained from scratch. Although the architectures are common choice in the state of the art and are available pre-trained, in this case it has been decided not to perform any fine-tuning technique that would involve the use of external data to those proposed in the dataset.

\subsection{Recurrent decoder}\label{subsec:decoder}

\par The decoder is implemented via a Recurrent Neural Network (RNN). As explained before, this block task is to look at the encoder's output and generate the final caption word by word. The decoder is implemented with a LSTM layer.  The final layer of the decoder corresponds to a Dense layer of 4637 (4635 words + \texttt{<eos>} + \texttt{sos>}) units with sigmoid activation that indicates the probability of each word to be used as caption. This decoder also implement an attention mechanism to allow the decoder to analyze different parts of the encoder's output as it has to predict the next word. In practice, this is a smart weighted average across all encoder's outputs feature maps. The higher weights will indicated relevancy in that features. This averaged feature map is passed as input to generate the next word. The attention mechanism takes into account the sequence created at the specific moment before predicting the next word and looks for the next part of the encoder's output to be more relevant. This is done by a a DNN with a softmax activation at the last layer in order to assign the relevancy of each part of the encoder's output. 

\subsection{Beam Search}\label{subsec:beam_search}

Different approaches are available in order to implement the decoding stage. The simplest way is to decode the highest scored word (the most likely one) at each iteration and consider it the best output caption. Although this approach is valid, it is not the optimal decoding process. At each iteration, the best possible word depends on the former selected words but it is possible that other combinations, considering not only the first best word but the $k$ best words, might be a better solution. As a result, $k$ sequences are obtained instead of only one and then, the best one is considered as the decoder output. This idea is called Beam Search. This method is used in the testing subset (see subsection~\ref{subsec:clotho}.)

In this case, beam width is 3, that is $k=3$. At first iteration, the 3 best words are considered and at second iteration, each one considers their 3 best options as a second word. At this point, there are 9 possible paths with their corresponding 9 combined scores. The 6 worst paths are pruned, keeping the other 3 at the end of the iteration. This procedure is repeated iteratively until each path reaches the \texttt{<eos>} token. When there are 3 completed sentences, their combined scores are calculated and the sentence with the best one is considered the output of the decoder.

\begin{table*}[]
\centering
\scalebox{0.85}{
\begin{tabular}{cccccccccc}
\toprule 
\textbf{System}         & \textbf{BLEU\textsubscript{1}} & \textbf{BLEU\textsubscript{2}} & \textbf{BLEU\textsubscript{3}} & \textbf{BLEU\textsubscript{4}} &
\textbf{METEOR} &
\textbf{ROUGE\textsubscript{L}} & \textbf{Cider}  & \textbf{SPICE}  & \textbf{SPIDEr} \\
\toprule
Baseline       & 0.344   & 0.082   & 0.023   & 0.000 & 0.066  & 0.234                                   & 0.022  & 0.013  & 0.018  \\ \midrule
Enc-50-DecAtt       & 0.464   & 0.260   & 0.157   & 0.092  & 0.135 & 0.308                                   & 0.195  & 0.083  & 0.139  \\ \midrule
\textbf{Enc-101-DecAtt} & 0.469  & 0.265  & 0.162  & 0.096 & 0.136 & 0.310                                  & 0.214 & 0.086 & 0.150 \\ \midrule
Enc-152-DecAtt &  0.466       &   0.261      &    0.156     &   0.091    & 0.137         &                          0.310     &     0.207    &    0.086    &   0.147    \\ \midrule
   Enc-Wide101-DecAtt     &    0.464     &       0.259   &   0.154      &  0.086        & 0.137 &              0.310  & 0.205     & 0.087   &  0.146     \\ \bottomrule
\end{tabular}}
\caption{Results with different Residual configurations plus Beam Search method in challenge results. Metrics obtained from testing subset.}
\label{tab:results_eval}
\end{table*}

\section{Experimental details}\label{sec:exp}

\subsection{Clotho dataset}\label{subsec:clotho}

\par In order to validate automated audio captioning systems the Clotho dataset will be used \cite{Drossos_2020_icassp}. This is the first captioning dataset manually labeled using only audio data information. All the information on how it has been labeled and the postprocessing procedure can be found in \cite{Drossos_2019_dcase}. This dataset is made up of audios from 15 seconds long to 30 seconds long. Each audio has 5 captions that can vary from 8 to 20 words. There are a total of 4981 audio samples and therefore 24905 captions. 

% All audios are from the Freesound platform and their titles have been made using the Amazon Mechanical Turk tool with annotators from English-speaking countries.

\par The total number of possible words in the dataset is 4365. In turn, the dataset is divided into 3 parts: development, evaluation and testing. Some of the details of this separation are that only an audio sample can appear in one of the partitions and not in two. There is no word that appears only in a split. The appearance of the words is proportional to the partition percentage: 60 \% development (training), 20 \% evaluation (validate) and 20 \% testing (challenge classification).

\par During the development phase of the contest, the development and evaluation partitions have been released. The development split consists of 2893 audio clips and 14465 captions. This partition is used to train the system. The evaluation split consists of 1045 audio clips and 5225 captions. This partition is used to validate the training and check its generalization. This block is used to choose the model that will be used to predict once the testing partition is released. All partitions take into account the consideration of the proportional appearance of words in each split as explained before.

\subsection{Training procedure}\label{subsec:training}

\par The system is optimized during training using Adam optimizer \cite{kingma2014adam}. The batch size is set to 16 samples and the number of epochs to 300. The loss function is set to cross-entropy loss. The learning rate start with a value of $10^{-4}$ and before every weight update, the 2-norm of the gradients clipped using as a threshold the value of 2. The systems have only been trained with the development subset. Captions are calculated with teacher forcing method \cite{lamb2016professor}.

\section{Results}\label{sec:results}

\subsection{Baseline system}\label{subsec:baseline}

\par The starting point of this task consists in an encoder-decoder structure formed by a Recurrent Neural Network (RNN) in both blocks. The encoder corresponds to a RNN of 3 bi-directional GRU layers. All GRU layers have the same number of features, 256.  The decoder in implemented with just one GRU layer of 256 features and a classification layer of 4637 (length necessary to represent all possible words in one-hot encoding).

\par The input is a log-Mel Spectrogram audio representation with 64 Mel filters. The temporal bins are obtained using a 46 ms window length with 50\% overlap. Representation are padded with zeros at the beginning when the audio length is less than the maximum possible in the batch.

\subsection{Metrics}\label{subsec:metrics}

There are a series of metrics that allow evaluating the performance of an autonomous captioning system. These metrics are: BLEU\textsubscript{$n$} \cite{papineni2002bleu}, ROUGE\textsubscript{L} \cite{lin-2004-rouge}, METEOR \cite{lavie-agarwal-2007-meteor}, CIDERr \cite{7299087}, SPICE \cite{anderson2016spice} and SPIDEr \cite{spider}. BLEU\textsubscript{$n$} calculates precision for $n$-grams, for that purpose, the number of matchings between the target caption and the predicted caption. BLEU\textsubscript{$n$} penalizes if the predicted caption is shorter than the target.BLEU\textsubscript{$n$} does not take into account the context. The subscript $n$ represent the grams. METEOR is considered a recall-based metric. The harmonic mean of precision and recall of segments of the captions between the predicted and the target ones is computed. It employs alignment between predicted and target words. This alignment is computed over segments of the caption, the number of chunks needed is the least possible. ROUGE\textsubscript{L} is a Longest Common Subsequence (LCS) based metric. The F-measure using LCS between the predicted and the target caption is computed. In this case, more weight is given to the recall \cite{Drossos_2017_waspaa}. CIDEr computes a weighted sum of the cosine similarity between both captions (predicted and target). Finally, SPICE is a more sophisticated metric that aims at capturing human judgments over the predicted caption better than the presented metrics. SPICE computation is divided into 3 steps. The first one consists in transforming both predicted and target captions into an intermediate representation that encodes semantic propositional content. This metrics focuses on semantic meaning. After this parsing a F-score caltulation is run. Finally, a gameability step is implemented in order to measure how well the predicted caption has recover objects, attributes and the relation between them. Take into account that this metric is an image-based metric. SPIDEr can be understood as a combination of SPICE and CIDEr metrics.

\subsection{Performance Analysis}\label{subsec:performance}

% \textcolor{blue}{https://arxiv.org/pdf/2006.03391.pdf ejemplo por si se quiere añadir algo mas}

Table~\ref{tab:results} shows the results obtained for the different implementations submitted to this task. As it can be appreciated, the study module in this task has been the encoder. For this, different state-of-the-art architectures incorporating residual learning have been implemented. 

Table~\ref{tab:results} presents all the metrics obtained on the evaluation subset. As it can be noticed, not all the metrics show an improvement of the baseline system. If the focus is on the SPIDEr metric, (since it is the one that is going to be used to rank the systems in the task) it can be noticed that the network with \textit{Resnet101} encoder shows a better behavior. The system obtains a value of 0.063, improving the result presented in the baseline. As it can be observed, all other settings show the same value in the SPIDEr metric. As it can be observed, with this subset of the dataset, the behavior of the rest of the implemented encoders is almost the same for the SPICE and CIDEr metrics. This can be due to the fact that the deeper networks to \textit{Resnet101} are suffering from overfitting. On the other hand, \textit{Resnet50} is too tiny to extract relevant features from the input Gammatone spectrogram.

% the system that implements the \textit{Resnet152} network shows a worse result, which may be a case of overfitting. On the other hand, \textit{Resnet50} shows the worst result, being in this case an example that the system is too tiny to extract relevant features from the input Gammatone spectrogram. On the other hand, the encoder with wide residual learning does not achieve any improvement in the classic residual architecture. It should be noted that all the metrics indicate the same situation. That is, \textit{Resnet101} achieves the best results in all metrics.

\subsection{Evaluation examples}\label{subsec:eval}

In this section some examples of how the proposed system generates captions from the provided audio are shown. The captions shown correspond to samples of the testing partition of the Clotho dataset.

\begin{table}[H]
\centering
\scalebox{0.88}{
\begin{tabular}{c|c}
\toprule
\textbf{Encoder}     & \textbf{Predicted Caption}                                                     \\
\toprule
Enc-50      & a car engine revs up and then drives away                   \\
Enc-101     & a car engine is revving up and is driving away              \\
Enc-152     & a motor is running and people are talking in the background \\
Enc-Wide101 & cars drive by on a busy road as people are talking  \\ \bottomrule       
\end{tabular}
}
\caption{Predicted captions for test\_0002.wav}
\label{tab:test2}
\end{table}

\begin{table}[H]
\centering
\scalebox{0.70}{
\begin{tabular}{c|c}
\toprule
\textbf{Encoder}     & \textbf{Predicted Caption}                                                     \\
\toprule
Enc-50      & people talk to each other while cars drive by                 \\
Enc-101     & a group of people are talking in the background    \\
Enc-152     & a crowd of people are talking in the background and gets louder as time goes on \\
Enc-Wide101 & people talk to each other while the vehicles drive by in the background  \\ \bottomrule       
\end{tabular}
}
\caption{Predicted captions for test\_0014.wav}
\label{tab:test14}
\end{table}

\begin{table}[H]
\centering
\scalebox{0.80}{
\begin{tabular}{c|c}
\toprule
\textbf{Encoder}     & \textbf{Predicted Caption}                                                     \\
\toprule
Enc-50      & water is running from a faucet and is filling up something                 \\
Enc-101     & water is running and dripping out of a faucet    \\
Enc-152     & water is running and a person is eating a shower \\
Enc-Wide101 & water is being poured from a pitcher into a cup and glass again  \\ \bottomrule       
\end{tabular}
}
\caption{Predicted captions for test\_0028.wav}
\label{tab:test28}
\end{table}

% \textcolor{blue}{Siempre se pueden poner mas ejemplos}

As it can be seen, the captions have an acceptable grammatical structure. Some of the visible differences may be the decision between the use of the present or the present continuous as can be seen in Table~\ref{tab:test2}. Another difference may be the decision when defining a concept as, for example, ``a group of people",  ``people" or ``a crowd of people" as can be seen in Table~\ref{tab:test14}. Table~\ref{tab:test28} is another example and how the system still generates nonsense ``eating a shower".

% For more clarity, a table is presented with the relationship between the system studied and the label used in the challenge submission 

% \begin{table}[H]
% \centering
% \begin{tabular}{cc}
% \toprule
% \textbf{Encoder used} & \textbf{Submission name}      \\
% \toprule
% Enc-50       & Naranjo-Alcazar\_UV\_task6\_1 \\ \midrule
% Enc-101  & Naranjo-Alcazar\_UV\_task6\_2 \\ \midrule
% Enc-152         & Naranjo-Alcazar\_UV\_task6\_3  \\ \midrule
% Enc-Wide101   & Naranjo-Alcazar\_UV\_task6\_4 \\ \bottomrule                             
% \end{tabular}
% \caption{Relationship between the name of the submission and the implementation explained in this paper.}
% \label{tab:submission}
% \end{table}

\subsection{Evaluation discussion}\label{subsec:discussion}

The systems with encoder \textit{Resnet50}, \textit{Resnet101}, \textit{Resnet152} and \textit{Wide101} have obtained a SPIDEr metric value of 0.139, 0.150, 0.147 and 0.146 respectively (see Table~\ref{tab:results_eval}).  As it can be observed it has been a huge improvement to the one shown in the development phase (see Table~\ref{tab:results}). The best system has ranked in the position 13 of 31 systems. As it can be appreciated, this system (\textit{Resnet101}) shows the best performance in all metrics.  If the rest of the encoders are analyzed, it can seen that \textit{Resnet50} shows a better behavior in the BLUE\textsubscript{$n$} metrics and \textit{Resnet152} and \textit{Wide101} in the rest of the metrics. The systems ranked between the first and eleventh positions have implemented different data augmentation techniques.  The system that ranks (without data augmentation) above those presented in this paper, has the difference of using embeddings when representing the words and implementing a CRNN as an encoder. The use of embeddings has also been used by the winning system. It has not been our case since it was not intended to add more degrees of freedom to the system, for example, the Word2Vec model and only analyze the behavior of the encoder-decoder.

\section{Conclusion}\label{sec:conclusion}

Audio captioning is a very novel task in the field of machine listening. Automated captioning is a problem that has been getting the attention of the image research community for a few years now. Thanks to the recent release of a dataset specially designed for audio captioning and the proposal of this task, the first novel solutions will be proposed. In this work, a state-of-the-art image captioning network (decoder-encoder structure) is implemented in the problem of audio captioning by making a study in the encoder block that is in charge of extracting the information from the audio. It has been decided to change the state-of-the-art representation based on Mel filters and to use the Gammatone filter bank that has shown good results in other machine listening tasks. The results presented in this paper show a considerable improvement of the baseline system in the chalenge results, enhancing its performance in the SPIDEr metric. In addition, the systems have shown competitive behaviour in the test phase in accordance with the limitations imposed on their design, such as the non-use of data augmentation techniques or word embeddings.

% -------------------------------------------------------------------------
% Either list references using the bibliography style file IEEEtran.bst
\bibliographystyle{IEEEtran}
\bibliography{refs}

%
% or list them by yourself
% \begin{thebibliography}{9}
% 
% \bibitem{dcase2016web}
%   \url{http://www.cs.tut.fi/sgn/arg/dcase2016/}.
%
% \bibitem{IEEEPDFSpec}
%   {PDF} specification for {IEEE} {X}plore$^{\textregistered}$,
%   \url{http://www.ieee.org/portal/cms_docs/pubs/confstandards/pdfs/IEEE-PDF-SpecV401.pdf}.
%
% \bibitem{PDFOpenSourceTools}
%   Creating high resolution {PDF} files for book production with 
%   open source tools, 
%   \url{http://www.grassbook.org/neteler/highres_pdf.html}.
%
% \bibitem{eWilliams1999}
% E. Williams, \emph{Fourier Acoustics: Sound Radiation and Nearfield Acoustic
%   Holography}. London, UK: Academic Press, 1999.
% 
% \bibitem{ieeecopyright}
%   \url{http://www.ieee.org/web/publications/rights/copyrightmain.html}.
%
% \bibitem{cJones2003}
% C. Jones, A. Smith, and E. Roberts, ``A sample paper in conference
%   proceedings,'' in \emph{Proc. IEEE ICASSP}, vol. II, 2003, pp. 803--806.
% 
% \bibitem{aSmith2000}
% A. Smith, C. Jones, and E. Roberts, ``A sample paper in journals,'' 
%   \emph{IEEE Trans. Signal Process.}, vol. 62, pp. 291--294, Jan. 2000.
% 
% \end{thebibliography}

\end{sloppy}
\end{document}